# Negative hydrogen ion production from a nanoporous 12CaO・7Al2O3 (C12A7) electride surface

Mamiko Sasao[1*], Roba Moussaoui[2], Dmitry Kogut[2], James Ellis[3], Gilles Cartry[2], Motoi Wada[4], Katsuyoshi Tsumori,[5] and Hideo Hosono[6]

[1] *Office for R&D promotion, Doshisha University, Kamigyoku, Kyoto, 602-8580, Japan*
[2] *Aix-Marseille Univ., CNRS, PIIM, UMR 7345, 13397 Marseille, France*
[3] *York Plasma Institute, Department of Physics, University of York, York YO10 5DD, United Kingdom*
[4] *Graduate School of Science and Engineering, Doshisha University, Kyotanabe, Kyoto, 610-0321, Japan*
[5] *National Institute for Fusion Science, 322-6 Oroshi, Toki, Gifu 509-5292, Japan*
[6] *Material Research Center for Element Strategy and Materials and Structures, Laboratory, Tokyo Institute of Technology, Yokohama, 226-8503, Japan*

E-mail: msasao@mail.doshisha.ac.jp

**Abstract**

A high production rate of negative hydrogen ion was observed from a nanoporous C12A7 electride surface immersed in hydrogen/deuterium low-pressure plasmas. The target was biased at 10 – 170 V negatively and the target surface was bombarded by $H_3^+$ ions from the plasma. The production rate was 1-2 orders higher than a clean molybdenum surface. The measured $H^-$ energy spectrum indicates that the major production mechanism is desorption by sputtering. This material has potential to be used as production surface of cesium-free negative ion sources for accelerators, heating beams in nuclear fusion, and surface modification for industrial applications.

Large-scale negative ion beams of hydrogen isotopes have been intensively developed as plasma heating and current-drive tools in nuclear fusion projects[1,2]. Medium-size negative ion beams are widely used nowadays as injectors for particle accelerators, such as those of J-PARC, CERN, SNS-Oak Ridge, China Spallation Neutron Source (CSNS), Los Alamos Neutron Science Center (LANSCE)/PFR, ISIS Spallation Neutron and Muon Facility, and so on[3]. Small-size negative ion sources are now under development for the application to accelerators in medical area[4]. Negative ion sources have recently received attention due to their application in microelectronics areas, such as, proton implantation for surface modification, or smart-ion-beam cut of wafers[5,6], because the charge-up problem of positive ion injection can be avoided.

It is known that H$^-$ ions can be produced in a plasma volume[7], and on a low-work-function surface covered by a half or less monolayer of alkali metals[8,9]. In most of large- or medium-size ion sources, cesium vapor is introduced not only for an efficient H$^-$ surface production but also for the remarkable reduction of co-extracted electron current. However, it has many disadvantages, such as limited life time of the coverage, cesium leakage. Cesium contamination on surfaces to be treated, and surroundings is a fatal drawback for application to microelectronics. Research is currently being carried out in order to develop cesium-free negative-ion sources. Alternative materials, such as carbon materials, specially highly oriented pyrolitic graphite (HOPG) or diamond, are studied to enhance the negative ion production on a plasma grid of an ion sources[10,11].

In 2003 Hosono and his coworkers have found that a refractory oxide 12CaO.7Al2O3 (C12A7) can be transformed to an electride of low work function[12,13,14], by exchanging of clathrated oxygen ions as the counter anion from the positively charged crystallographic cages of C12A7 leading to formations of high-density electrons in the cages. When 100% of clathrated oxygens were removed, it showed conductivity approximately 1000 siemens per centimeter at 300 K, demonstrating that the encaged electron behaves as an anion[15]. They reported that the resulting [Ca24Al28O64]$^{4+}$(4e-) may be regarded as an "electride", and clarified the unique properties,i.e., they have a very low work function of 2.4 eV, which is comparable to that of metal potassium, but chemically and thermally stable. Figure 1 shows the schematic energy diagram of C12A7 electride. In our previous work, H$^-$ formation from a sample of C12A7 electride, was indirectly shown by exposing the sample to an atomic hydrogen (H$^0$) flux and detecting the negative current generated on it[16]. The current observed was almost similar level with that obtained from a low-work-function bi-alkali covered molybdenum surface (~2.3 eV) in the same condition.

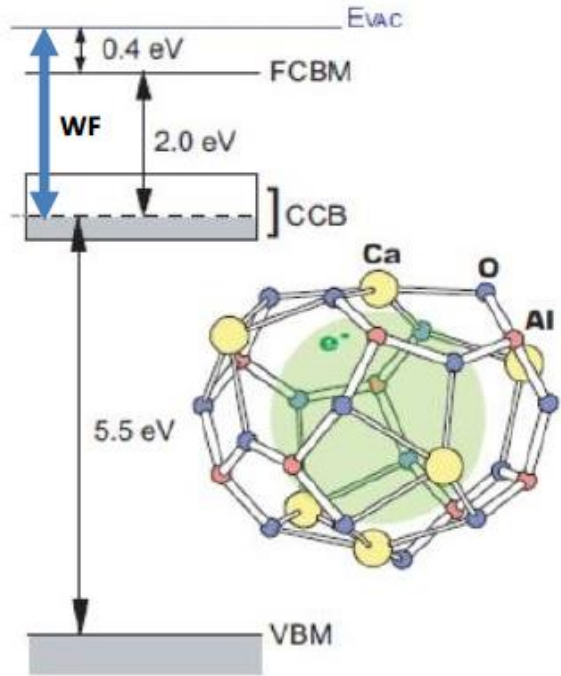

FIG. 1. Energy diagram of C12A7 electride. Anionic electrons are accommodated in the sub nanometer-sized cages which are connected with each other. The connected cages form a new conduction band called "cage conduction band" (CCB), below the cage flame conduction band minimum (FCBM). The work function (WF) is 2.4 eV, comparable to metal potassium, but chemically and thermally stable owing to that electrons are sitting within the cages composed of rigid Ca-AL-O network. $E_{VAC}$: vacuum level, VBM: valence band maximum. The band gap is 7.5eV.

In the present work, direct measurement of H⁻ formation was carried out by an experimental apparatus in Aix-Marseille University. It consists of an upper plasma source chamber and a lower spherical diffusion chamber. The plasma was initiated by a Huttinger PFG 1600 RF generator followed by a Huttinger matchbox. A sample, which was able to be retrieved into a separate vacuum chamber for change, was fixed on a rotatable target holder and inserted into the center of the lower diffusion chamber. The temperature of the target holder and the target itself was controllable from room temperature up to 600° C, with heating arrangement. An ECR plasma source was also equipped, and a plasma could be injected in the lower chamber towards the target. The target was biased negatively against the chamber. Fig. 2 shows the electrical potential profile between the target sample holder and the mass spectrometer. Due to the electrical potential drop, $e(V_p - V_s)$ in front of the target sample, positive ions were accelerated and bombarded the surface. Here $V_p$ and $V_s$ denote the plasma potential and the bias voltage on the target sample, respectively. Most of ions were injected into the sample or stop near the surface, but some of them were reflected on it. However due to the potential rise, only neutral particles, reflected negative ions and particles sputtered out as negative ions could escape from the surface. Negative ions were accelerated by $e(V_p - V_s)$ and reached the sheath region in front of the grounded entrance electrodes of the mass spectrometer, and decelerated by $e(V_p)$. Consequently, the total energy gain of negative ions was $-eV_s$. In the present experiment, negative ions were energy analyzed by a Hiden EQP300 mass spectrometer equipped with an energy filter.[17] The aperture plate of it was placed 36 mm apart from the target surface, with the center of sample located on the axis of the spectrometer. Details on the experimental set-up and experimental conditions are given in the reference[18].

The measurement shown here were made at 2 Pa hydrogen or deuterium gas pressure fixed, from 20 to 250 W injected power for RF plasma conditions, and 1 Pa 60 W injected power for ECR conditions. In all plasmas the dominant ion species was $H_3^+$. The target was kept at room temperature during the measurement. Plasma parameters were measured by a Langmuir probe in a separate experiment under the same condition. Typical values were ranging in $T_e \sim 1$ eV, $n_e \sim 10^9/cm^3$ for the ECR operation, and $T_e \sim 3$ eV, $n_e \sim 10^8/cm^3$ for the RF operation when excitation of coils for vertical magnetic field was turned off in order to avoid the interference with the spectrometer. Plasma potential, $V_p$, was about 8 V for both cases. Major negative ion impurities were OH$^-$ and H$^-$. During the present experiment, the spectra of these impurities were measured occasionally and their peak values were about 3 % and 0.2 % of that of H$^-$, respectively.

The negative ion energy distribution function (NIDEF) shown in FIG. 3a and FIG. 3b were measured by the energy-analyzed mass spectrometer from C12A7 electride (a) and a molybdenum target (b) with $D_2$ plasmas of RF of 250 W. Imprity negative ions the vertical axis is counts per second detected by the mass spectrometer, and horizontal axis is the negative ion energy after subtraction of energy gain by the target bias voltage, so that it corresponds to the initial energy emitted from the target. Two lines shown in the figures are spectra with the target voltage of -20 V (red) and -80 V(black). The target was placed normal to the spectrometer axis, and most of particles detected were emitted almost normally. The effects of angular distribution of ions from a target, their trajectories, and collection probability by the spectrometer on the measured spectra were studied by simulations and discussed in the references[10,11]. A large peak in the low energy region less than 10 eV was seen only in the C12A7 electride spectra. It has been known that low energy negative hydrogen ions are produced through the volume production, where vibrational excited $H_2$ (X,v) molecules are dissociated into $H^-$ and $H^0$ in conjunction with electron attachment[19]. It was found that some metallic surface revealed an enhancing effect on highly vibrationally excited hydrogen molecules $H_2(X,v)$ in gas phase experiments[20]. However, the present potential profile shown in FIG. 2 prevented negative ions produced in the plasma from coming into the sheath region in front of the analyzer.

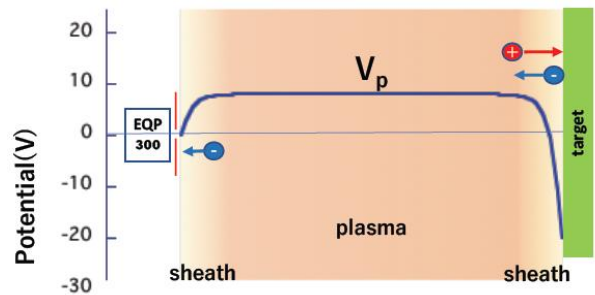

FIG. 2  Electrical potential profile between the sample holder and the mass spectrometer EQP 300, when the target sample was biased at – 20 V. Horizontal axis is not scaled.

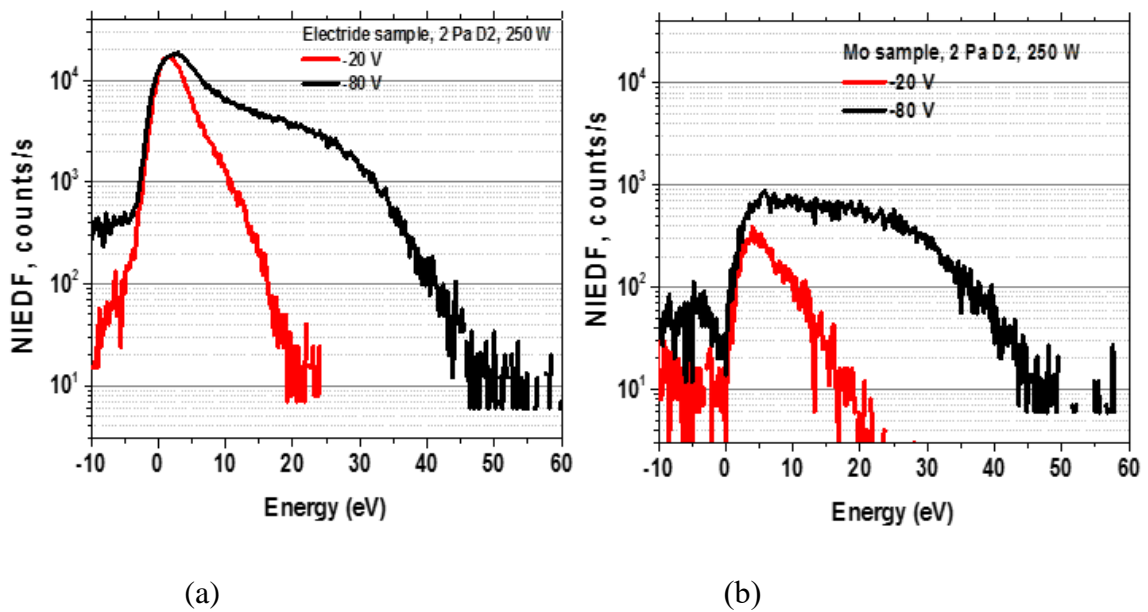

(a)          (b)

FIG. 2.  Negative ion spectrum from a C12A7 elctride (a) and molybdenum (b) in RF plasma. The red and black lines show those obtained when the target bias was -20 V and -80 V, respectively.

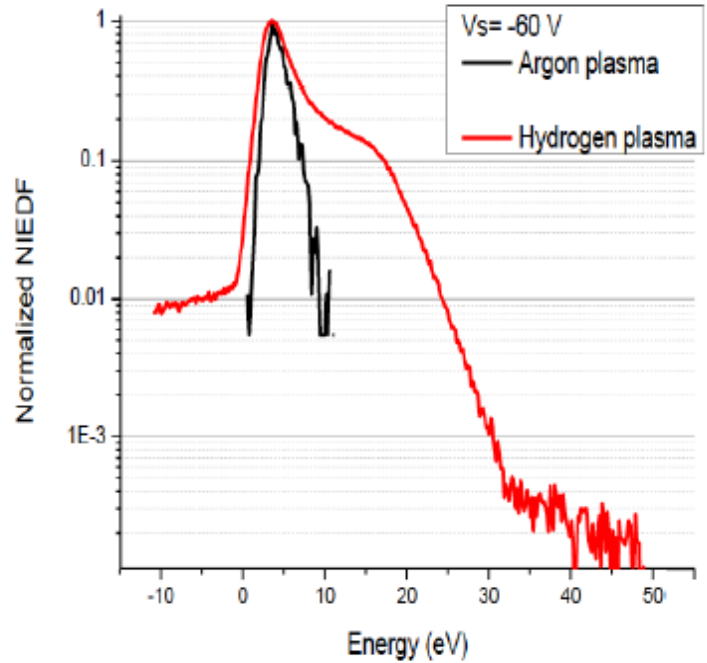

FIG. 3. Comparison of normalized negative ion energy spectra measured in an ECR Ar plasma (black) under 0.1 Pa, 30 W, and in an ECR $H_2$ plasma (red) under 1 Pa, 60 W at $V_S = -60$ V after the sample has been treated for 10 minutes in an ECR $H_2$ plasma (1 Pa, 60 W, $V_S = -130$ V).

In order to confirm that the negative ions from the C12A7 electride was produced by desorption in which implanted H/D particles were sputtered out, two measurement were carried out. In the first experiment, energy spectra of hydrogen negative ions from the C12A7 electride bombarded by hydrogen and by argon ions were compared as shown in FIG.3 In both measurement, the target was pre-treated in the same procedure with an exposure to hydrogen ECR plasma (1 Pa $H_2$, 60 W) at $V_S= -130$ V for 10 min. Then, hydrogen plasma at 1 Pa, 60 W, $V_S = -60$ V or argon plasma at 0.1 PA, 30 W, $V_S = -60$ V were switched on and negative-ion spectrum were recorded. It turned out that the dominant peak shown in the low energy region below 10 eV was those due to the sputtering of pre-implanted hydrogen particles.

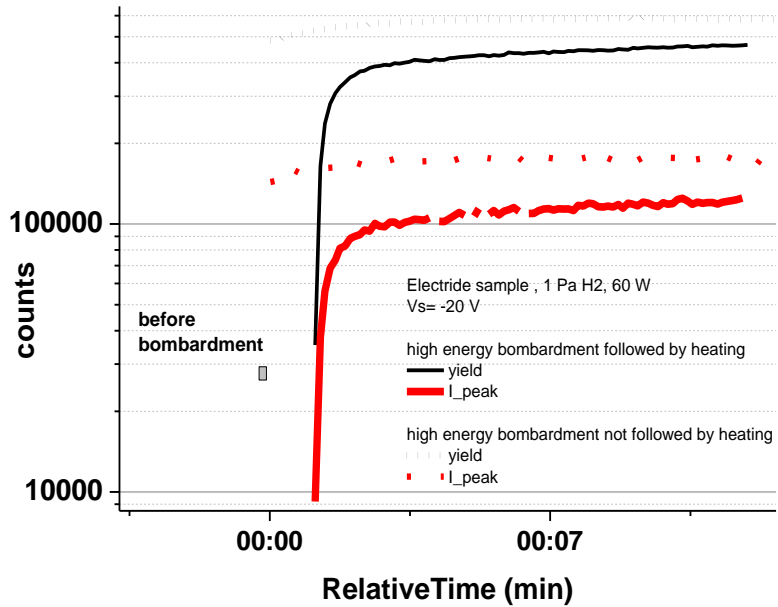

FIG. 4. Time evolution of Negative ion yield after 10 min. of high energy bombardment followed by a heating in vacuum and not followed by a heating in vacuum.

In the second experiment, the variation of hydrogen negative ion signals after baking of the sample was measured. Before the baking, several hours of biasing experiment in an ECR plasma of 60 W had been carried out (1 Pa, 60 W, $V_S = -130$ V). Then, 520° C baking was applied in vacuum for 50 min. During heating, a mass scan of hydrogen H2 and H2O variation was recorded, and the amount of hydrogen increased 3 times during heating and the amount of H2O increase 24 times during heating. After cooling down, an ECR hydrogen plasma was ignited (1 Pa, 60 W, $V_S = -20$ V). In FIG 4, the time behaviors of total negative ion produced and peak current of the spectra right after the application of – 20 V bias voltage are shown by black and red lines, respectively. The negative ion peak intensity increased by a factor of 11.7 the first 5 min, and it increased then slowly. For comparison purpose Figure 4 also presents the time behaviors of total negative ion produced and peak current of the spectra right after the application of – 20 V bias voltage without any baking. In this case, the sample has been exposed to a long ECR plasma exposure at 1 Pa, 60 W, $V_S = -130$ V before the measurements at $V_s = -20$ V. We can see that in this case the peak intensity increases only by a factor 1.7 instead of 11.7. It clearly shows that hydrogenation of the material and negative-ion creation through sputtering is explaining the low-energy part of the spectrum.

In these experiment, high negative hydrogen ion ($H^-$) production rate from a C12A7 electride target surface was observed, and there were negative ions produced through (1) the charge exchange (electron pick-up) reflection process and (2) the desorption process by sputtering[18]. The latter comportment was remarkably high compared with those from molybdenum. It has been known that negative ion production ratio strongly depends on the surface work function in both processes. The high production yield of this material may due to its low work function of 2.4- 2.7 eV[16], but also due to the specific nano-structure of C12A7 electride. When this electride is exposed to hydrogen circumstance, some of the encaged electron are replaced by the hydrogen negative ions. Although the nano-crystal structure is not preserved in the as-cleaved surface, the fractured cages can be easily restored upon heating at appropriate conditions[19].

In the present work, absolute production rate of negative hydrogen ion ($H^-$) was not obtained. However, electric current corresponding to negative hydrogen ion current observed by the exposure to the C12A7 electride to an atomic hydrogen ($H^0$) flux was almost similar level to that obtained from a low-work-function bi-alkali covered molybdenum surface (~2.3 eV) in our previous indirect measurement[20]. The dependence on the atomic hydrogen ($H^0$) power and other characteristics were also similar. Together with the present direct observation of the negative hydrogen ions, it is likely that this material has a potential to produce similar level of negative ions with the cesiated molybdenum surface.

This new material, C12A7 electride is air-stable, mechanically robust and machinable, and has potential to be used as production surface of cesium-free negative ion sources for accelerators, plasma heating beams in nuclear fusion, and surface modification for industrial applications.

ACKNOWLEDGMENTS

The C12A7-Electride was supplied from Asahi Glass Co. LTD., and technical support by Naomichi Miyakawa, Satoru Watanabe, and Kazuhiro Ito of Asahi Glass Co. LTD and deep discussion with them are greatly acknowledged. The authors would like to T. Yokoyama for the use of C12A7-Electride. HH acknowledges a fund from JST ACCEL program and JSPS KAKENHI Grant Number JP 17H061253). JE acknowledges a support from the CDT-Fusion and the EPSRC (No. EP/L01663X/1).

The experimental set-up at PIIM has been developed thanks to the financial support of the French Research Federation for Fusion Studies (FR-FCM) and of the French Research Agency (ANR) under grant 13-BS09-0017 H INDEX TRIPLED. This work was performed under the collaboration program of National Institute for Fusion Science (NIFS15KBAR011 and NIFS15KLER041).

**REFERENCES**

1) R.S. Hemsworth, New J. Phys. **19.** 025005 (2017).
2) K. Tsumori, et al., Rev. Sci. Instrum. **87** 02B936-1-6 (2016) .
3) W. Alexander et al., Reviews of Accelerator Science and Technology: Volume **6**: Accelerators for High Intensity Beams (2014)
4) H. Etoh et al., "DEVELOPMENT OF A HIGH CURRENT NEGATIVE ION SOURCE FOR MEDICAL CYCLOTRONS ", Proceedings of the 13th Annual Meeting of Particle Accelerator Society of Japan (August 8-10, 2016, Chiba, Japan), PASJ2016 MOP052 (2016).
5) H.J. Woo, et al., "APPLICATION OF HYDROGEN ION BEAMS FOR SOI WAFER FORMATION", Proceedings of APAC 2004, Gyeongju, Korea, pp. 435 – 437 (2004).
6) S. Ahmed, et al, "Proton implantation for effective electrical isolation of InP, InGaAs and GaAs: Role of variable doses and implant temperature", Proceedings of 14th Indium Phosphide and Related Materials Conference (IEEE, 2002, Stockholm),.
7) M. Bacal and G. W. Hamilton, Phys. Rev. Lett. **42**, 1538 (1979).
8) M. Nishiura, M. Sasao and M. Wada, "Measurements of H- density and work function of a plasma electrode in a negative ion source", Proceedings of 9th Int. Symp. on Production and Neutralization of Negative Ions and Beams, AIP Conf. Proc. **639**, 21 (2002).
9) A. A. Padama, Journal of the Vacuum Society of Japan Vol. **57**, No. 1 (2014)
10) G. Cartry, et al., A, *New Journal of Physics* **19** 025010 (2017)
11) A. Ahmad, C. Pardanaud, M. Carrèr, J-M. Layet, A. Gicquel, P. Kumar, D. Eon, C. Jaoul, R. Engeln and G. Cartry, *Journal of Physics D: Applied Physics* **47** 085201 (2014).
12) H. Hosono, Japanese Journal of Applied Physics **52**, 090001 (2013).
13) Peter V. Susko, et al., Phys. Rev. Lett. **91**, 126401 (2003).9